\documentstyle[prl,aps,epsf,multicol]{revtex}
\newcommand{\be}{\begin{equation}}
\newcommand{\ee}{\end{equation}}
\newcommand{\bea}{\begin{eqnarray}}
\newcommand{\eea}{\end{eqnarray}}
\begin{document}

\title{Extreme Value Statistics and Traveling Fronts: 
An Application to Computer Science}
\author{Satya N. Majumdar$^{1}$ and P. L. Krapivsky$^2$}
\address{ 
 {\small $^1$Laboratoire de Physique Quantique (UMR C5626 du CNRS),
Universit\'e Paul Sabatier, 31062 Toulouse Cedex, France}\\
{\small $^2$Center for Polymer Studies and
Department of Physics, Boston University, Boston, MA 02215, USA}}
\maketitle
\widetext

\begin{abstract} 
  
  We study the statistics of height and balanced height in the binary search
  tree problem in computer science. The search tree problem is first mapped
  to a fragmentation problem which is then further mapped to a modified
  directed polymer problem on a Cayley tree. We employ the techniques of
  traveling fronts to solve the polymer problem and translate back to derive
  exact asymptotic properties in the original search tree problem. The second
  mapping allows us not only to re-derive the already known results for
  random binary trees but to obtain new exact results for search trees where
  the entries arrive according to an arbitrary distribution, not necessarily
  randomly. Besides it allows us to derive the asymptotic shape of the full
  probability distribution of height and not just its moments.  Our results
  are then generalized to $m$-ary search trees with arbitrary distribution.

\end{abstract}
\begin{multicols}{2}

\section{Introduction}

The techniques developed in statistical physics, particularly in the theory
of spin glasses, have been recently applied to a variety of problems in
theoretical computer science\cite{MMZ}. These include various optimization
problems such as the traveling salesman problem\cite{MP}, graph
partitioning\cite{FA}, satisfiability problems\cite{MZ}, the knapsack
problem\cite{Inoue}, the vertex covering problem\cite{WH}, error correcting
codes\cite{SM}, number partitioning problems\cite{Fu}, matching
problems\cite{MPA} and many others\cite{MPV}. The purpose of this paper is to
study analytically certain problems in a different area of theoretical
computer science known as sorting and searching\cite{Knuth}. The standard
techniques of spin glass theory are not directly suitable for these problems.
Instead, we employ the techniques developed to study the propagation of
traveling fronts in various nonlinear systems
\cite{Fisher,KPP,Bramson,Zel,Murray,VS1,VS2}.

The ``sorting and searching'' is an important area of computer science that
deals with the following basic question: How to organize or sort out the
incoming data so that the computer takes the minimum time to search for a
given data if required later?  Amongst various search algorithms, the binary
search turns out to be one of the most efficient\cite{Knuth}.  To understand
this algorithm, let us start with a simple example. Suppose the incoming data
string consists of the twelve months of the year appearing in the following
order: July, September, December, May, April, February, January, October,
November, March, June and August. Suppose later we need to look for the month
of August in this data string.  Consider first the sequential search where
the computer starts from the first element (July), checks if it is the right
month and if not, moves to the next element of the string (September), checks
the element there and continues in this fashion until it finds the right
month.  In the example above, to find the month `August', the computer has to
make 12 comparisons.  Thus, the sequential search algorithm is rather
inefficient as it typically takes a search time of order $N$, where $N$ is
the number of entries in the data string.

In a binary search, on the other hand, the typical search time scales as
$\log N$\cite{Knuth}.  The binary search is implemented by organizing the
data string on a tree according to the following algorithm.  An order is
first chosen for the incoming data, e.g., it can be alphabetical or
chronological (January, February, March, etc.).  Let us choose the
chronological order. Now the first element of the input string (July) is put
at the root of a tree (see Fig. 1).  The next element of the string is
September. One compares with the root element (July) and sees that September
is bigger than July (in chronological order). So one assigns September to a
daughter node of the root in the right branch. On the other hand, if the new
element ware less than the root, it would have gone to the daughter node of
the left branch. Then the next element is December. We compare at the root
(July) and decide that it has to go to the right, then we compare with the
existing right daughter node (September) and decide that December has to go
to the node which is the right daughter of September. The process continues
till all the elements are assigned their nodes on the tree. For the
particular data string in the above example, we finally get the unique tree
as shown in Fig. 1.  Such a tree is called a binary search tree (BST).

Once this tree is constructed, the subsequent search, say for the month of
August, takes much less number of comparisons. We start with the root (July).
Since the sought after element August is bigger than July, we know that it
must be on the right branch of the two daughter sub-trees. This already
eliminates searching roughly half of elements which are on the left
sub-tree.  We next encounter the key September.  Since August is less than
September, we go to the left and thus we do not need to search anymore
the right branch of the sub-tree rooted at September.  Once we go to the
left, we find the required key August.

\begin{figure}
  \narrowtext\centerline{\epsfxsize\columnwidth \epsfbox{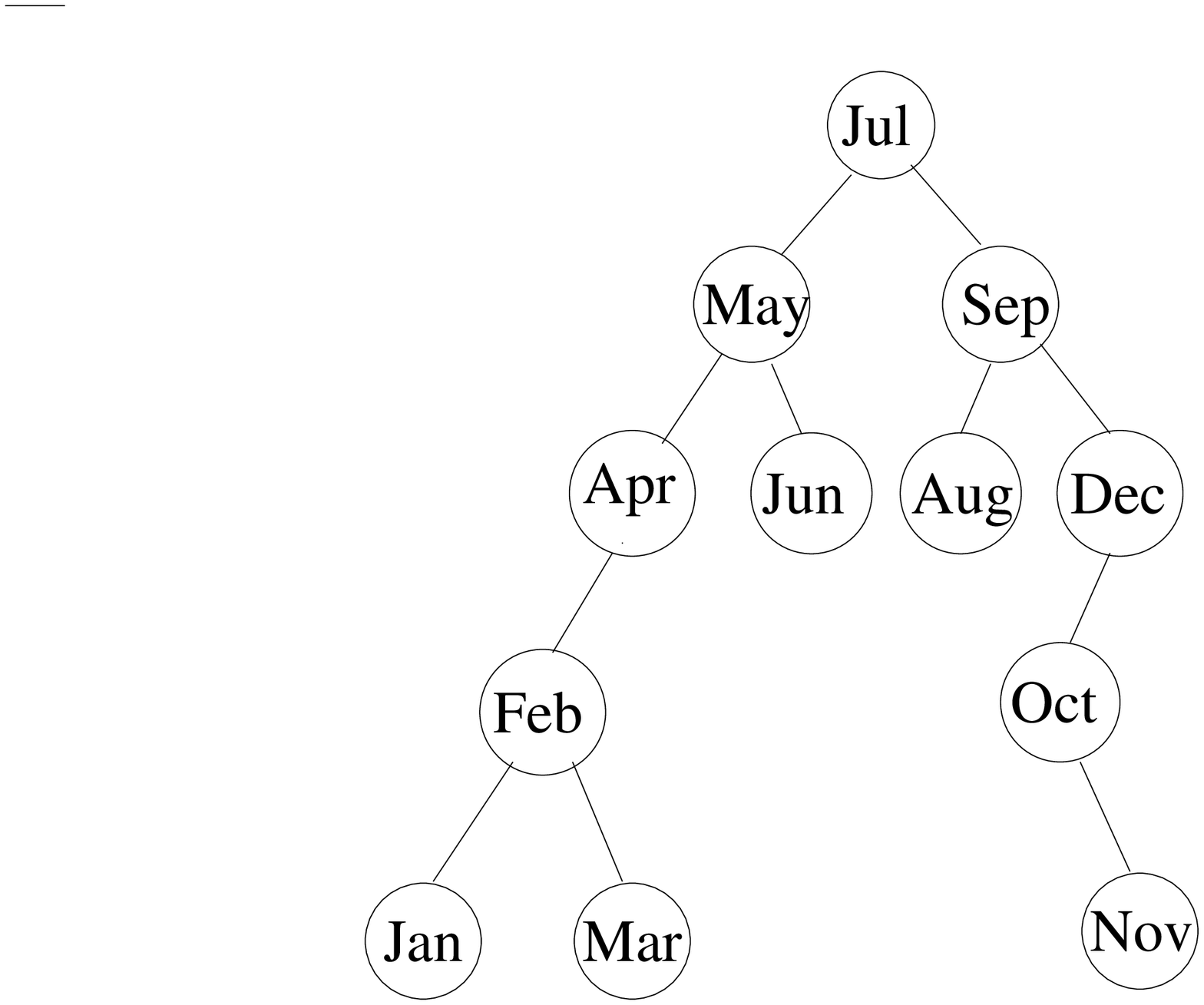}}
\caption{The binary search tree corresponding to the data string  
  in the order: July, September, December, May, April, February, January,
  October, November, March, June and August.}
\end{figure}
\noindent

Thus, the BST algorithm requires only 3 comparisons as opposed to 12
operations in the sequential search. Since the typical search time is
proportional to the depth of an element in the tree and since the typical
depth $D$ is related to the total size $N$ via $2^D\approx N$, the search
time scales as $\log N$, making the BST algorithm one of the most efficient
search algorithms.

\begin{figure}
  \narrowtext\centerline{\epsfxsize\columnwidth \epsfbox{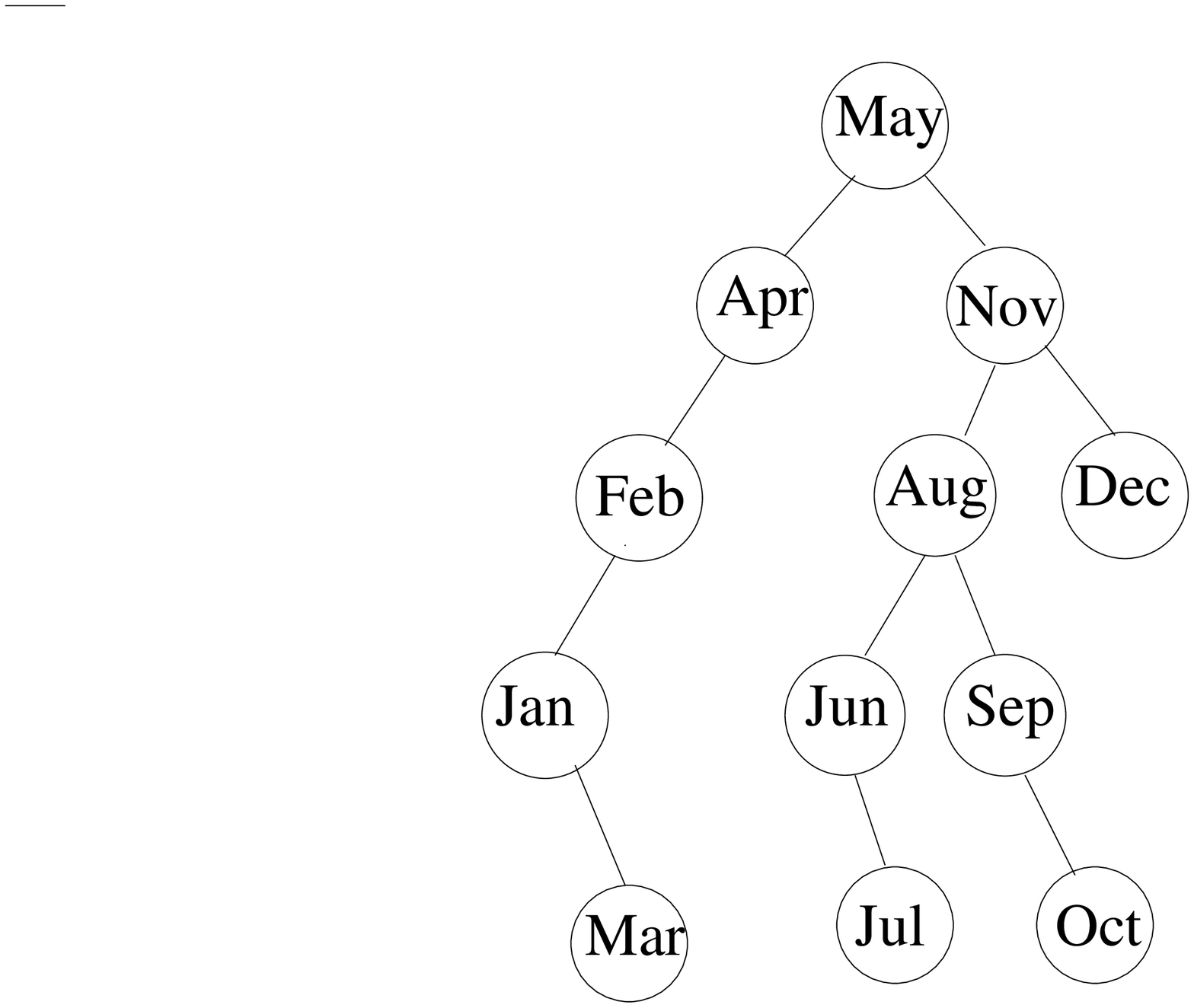}}
\caption{The binary search tree corresponding to the data string  
  in the order: May, November, August, April, December, February, June,
  September, July, January, October and March.}
\end{figure}

If the incoming data string had a different order of appearance, one would
have obtained a different BST. For example, suppose the months appear in a
different order as: May, November, August, April, December, February, June,
September, July, January, October and March. For this data string, the same
algorithm of constructing a binary tree as before gives a tree of different
shape (Fig.~2).  Each permutation of the incoming data string leads to a
different binary tree and there are $N!$ possible binary trees for any
incoming data string with $N$ entries. Usually the incoming data string
appears in a random fashion. This would indicate that each of the $N!$
possible binary trees occurs with equal probability. Such trees, each
generated with equal probability, are called `random binary search trees'
(RBST).  Of course, if the incoming data is not completely random, the
probability measure over the space of trees will not be uniform. The results
derived in this paper will be applicable not only to RBST but also to more
general BST's with arbitrary measure.

Each BST has several observables (such as the depth or the height of a tree)
associated with it that quantify the efficiency of the underlying search
algorithm. Hence the knowledge of the statistics of such observables are of
central importance.  Here are a few observables:

\vspace{\baselineskip}

$\bullet$ $D_N$ is the depth (distance from the root) of the last
inserted element in a given BST of size $N$.  For example, $D_N=3$ for the
tree in Fig.~1 (counting the depth of the root element as
$1$). Each BST has a different $D_N$, so $D_N$ is a random variable. The
average depth $\langle D_N\rangle$ (averaged over the probability measure of
the trees) gives a measure of the average time required to insert a new
element in a tree\cite{Knuth}.

\vspace{\baselineskip}

$\bullet$ $H_N$ is the height of a given tree, defined as the depth of the
farthest node from the root.  For example, $H_N=5$ for the BST in Fig.~1.
Clearly $H_N$ is also a random variable and $\langle H_N\rangle$ gives a
measure of the {\it maximum} possible time that could be required to insert
an element, i.e.  a measure of the {\it worst} case
scenario\cite{Robson,FO,D1,Mahmoud}.

\vspace{\baselineskip}

$\bullet$ $h_N$ is the balanced height defined as the {\it maximum} depth
from the root up to which the tree is {\it fully} balanced, i.e., all the
nodes up to this depth are fully occupied\cite{Knuth}.  In the tree shown in
Fig.~1, $h_N=3$, whereas $h_N=2$ in the tree in Fig.~2.  Hence $h_N$ is also
a random variable whose statistics is important.

Some of the observables mentioned above such as the height $H_N$ and the
balanced height $h_N$ are of extremal nature, i.e., they are the maximum or
the minimum of a set of {\it correlated} random variables. In this paper we
limit ourselves only to such extreme observables of the binary tree. While
the statistics of the extremum of a set of {\it uncorrelated} random
variables is well understood\cite{Gumbel,Galambos,Berman}, little is known
about the same for correlated variables\cite{BM}. However, in the present
problem the random variables are correlated in a special hierarchical way
which facilitates analysis. We will see that the extreme
variables in the BST problem satisfy nonlinear recursion relations that admit
traveling front solutions in some suitable variables.  A lot is known about
the speed and the shape of such fronts appearing in various nonlinear
systems\cite{Fisher,KPP,Bramson,Zel,Murray,VS1,VS2}.  Below, we will use
these techniques to study the statistics of extreme variables in the binary
tree problem.

Some of our results for the RBST were already known which we will mention as
we go along. However, the approach used here is quite different from those
used by the computer scientists. Computer scientists tend to establish upper
and lower bounds to the quantity of interest (typically the average value or
the variance of the observable) and then tighten the bounds\cite{D1}. If the
bounds coincide, one obtains an exact result\cite{D1}.  Our approach, on the
other hand, is a typical physicist's approach.  The methods we use may not
always be rigorous in the strict mathematical sense, but they lead to exact
asymptotic results in a physically transparent way. Moreover, our approach
allows us not only to reproduce already known asymptotics for the average
height and the average balanced height of the RBST, but also to obtain new
information about the variance and even the asymptotic shapes of the full
probability distributions. Besides, our method goes beyond the RBST and
yields exact results for trees generated with arbitrary distributions.

Our approach utilizes two exact mappings which can be summarized as follows.
Following Devroye\cite{D2}, we first map the RBST problem to a random
fragmentation problem where an object of initial length $N$ breaks randomly
into two fragments, each of which further breaks randomly into two parts, and
so on.  The fragmentation problem is interesting on its own right as it
appears in the context of various physical problems such as the energy
cascades in turbulence\cite{GEL}, rapture processes in earthquakes\cite{NG},
financial crashes in stock markets\cite{SJ}, and the stress propagation in
granular medium\cite{qmodel}.  Some of the extremal problems in the random
fragmentation problem were studied recently by Hattori and Ochiai\cite{HO}
and by us\cite{KM}. The method used in our previous short paper\cite{KM}
allowed us to obtain exact asymptotic results for the average of the maximal
piece of the $2^n$ fragments after $n$ iterations. The statistics of this
maximal piece is closely related to the statistics of the height in the RBST.
However, this method was not easy to extend to the cases beyond the random
fragmentation, i.e. when the break point is chosen from an arbitrary
distribution, not necessarily uniform. We will see later that a fragmentation
problem with a given break point distribution corresponds to a BST problem
where the incoming entries to the tree appear according to a specific
distribution and not just randomly.

In this paper, we show that the fragmentation problem, with arbitrary break
point distribution, can further be mapped onto a modified directed polymer
(MDP) problem on a Cayley tree.  The MDP problem differs from the
conventional directed polymer (DP) problem on a Cayley tree studied by
Derrida and Spohn\cite {DS} due to the presence of a special constraint in
the MDP.  Derrida and Spohn were mostly interested in the finite temperature
spin glass transition in the DP problem.  Our problem reduces to a zero
temperature problem, albeit with a special constraint. We then solve this MDP
problem using traveling front techniques and translate back to derive exact
asymptotic results for the original BST problem. We will see that the
statistics of the height $H_N$ of the BST problem is related (via the two
successive mappings) to the statistics of the minimum or the ground state
energy of the MDP problem.  On the other hand, the statistics of the balanced
height $h_N$ will be related to that of the maximum energy of the directed
polymer (a quantity of little interest in statistical physics framework).
This second mapping also allows us to obtain new exact results for nonrandom
BST problem.

The paper is organized as follows. In Sect.~I, we set up notation, review
known results for the RBST problem, and summarize novel results obtained in
this paper.  Section II contains the exact mapping of the BST problem to the
fragmentation problem. In Sect.~III, we map the fragmentation problem to the
MDP problem.  In Sect.~IV, we derive the exact nonlinear recursion relations
in the MDP problem and analyze them using the traveling front techniques. The
main results for the RBST problem are also derived in this section. In
Sect.~V, we go beyond the random trees and derive exact results for the
fragmentation problem with arbitrary break point distribution. Section VI
contains the generalization to the case of $m$-ary trees with arbitrary
distributions.  We finally conclude in Sect.~VII with a summary and outlook.

\section{Binary Search Trees: Old and New Results}

Let us label the incoming data string of $N$ elements by integers $1$,
$\ldots$, $N$.  For example, if the data string consists of the $12$ months
of the year, we can label, say the month of January by $1$, the month of
February by $2$, and so on.  In that example, $N=12$. A specific data string
will then be isomorphic to a corresponding ordered sequence of these
integers. For example, the particular sequence of months in Fig.~1 reduces to
the ordered sequence $[7,9,12,5,4,2,1,10,11,3,6,8]$. A different sequence of
months will correspond to a different permutation of these integers. Each
such sequence or permutation will then correspond to a separate BST,
constructed by the algorithm explained in the introduction. In RBST, all
these $N!$ sequences (and their corresponding trees) occur with equal
probability.

We will focus on the statistics of the extreme variables associated with
these trees, in particular the height $H_N$ and the balanced height $h_N$ of
a BST as defined in the introduction. Each BST has a unique value of $H_N$
and $h_N$.  Since the trees occur with a given probability distribution
(which is uniform in case of RBST), both $H_N$ and $h_N$ are random
variables. Of interest are the statistics of these variables such as the
average, variance or even the full probability distributions of $H_N$ and
$h_N$.

The RBST problem has been studied for a long time by computer scientists and
we now mention a few known results.  Devroye\cite{D1} proved that for large
$N$, the average height of a RBST $\langle H_N\rangle\approx \alpha_0 \log N$
where the constant $\alpha_0=4.31107\dots$. Hattori and Ochiai conjectured
that the true asymptotic behavior of $\langle H_N\rangle$ has an additional
sub-leading double logarithmic correction,
\begin{equation}
\langle H_N\rangle \approx \alpha_0 \log N +
\alpha_1 \log \left(\log N\right),
\label{HN1}
\end{equation} 
and they determined the constant $\alpha_1\approx -1.75$
numerically\cite{HO}.  Using traveling front techniques we confirmed the
above asymptotics and computed the correction term analytically,
$\alpha_1=-3\alpha_0/[2(\alpha_0-1)]=-1.95303\dots$ \cite{KM}.  The same
result was simultaneously proved by Reed\cite{Reed}. Based on numerical data,
Robson conjectured\cite{Robson2} that the variance is bounded.  Recently,
Drmota\cite{Drmota} has proved that all moments $\langle (H_N-\langle
H_N\rangle)^m \rangle $ are bounded.

For the balanced height $h_N$ of RBST, Devroye showed that the leading
asymptotic behavior of the average balanced height is given by $\langle
h_N\rangle \approx {\alpha_0}'\log N$ where
${\alpha_0}'=0.3733\dots$\cite{D1,D2}. Indeed, $\alpha_0$ in $\langle
H_N\rangle$ and ${\alpha_0}'$ in $\langle h_N\rangle$ turn out to be the two
solutions of the same transcendental equation $(2e/\alpha)^{\alpha}=e$
\cite{D1,D2}. This suggests some kind of duality between the height and the
balanced height. We will show later that the correct asymptotic behavior of
$\langle h_N\rangle$ is given by
\begin{equation} 
\langle h_N\rangle 
\approx {\alpha_0}' \log N + {\alpha_1}' \log \left(\log N\right),
\label{hN1}
\end{equation} 
where relation $\alpha_1'=-3\alpha_0'/[2(\alpha_0'-1)]$ holds again. Drmota
has recently proved that all the moments of $h_N$ are also
bounded\cite{Drmota} as in the case of $H_N$.

Note that all the results mentioned above are for RBST with fixed size $N$.
Recently by using a rate equation approach we studied the statistics of
height and balanced height for randomly growing binary trees where the
average size of a tree grows with time linearly $\langle N(t)\rangle \sim
t$\cite{BKM}.  The expected height and balanced height for large random
binary trees were found to have exactly the same asymptotic formulas
(\ref{HN1})--(\ref{hN1}), provided one replaces $N$ by $\langle N(t)\rangle$
in these equations. This approach is thus reminiscent of the grand canonical
approach in statistical mechanics with the time $t$ playing the role of the
chemical potential which can be chosen to fix the average size. In this
paper, we focus only on the canonical approach, i.e., trees with fixed given
size $N$, since this is more familiar in theoretical computer science.

We will exploit a two stage mapping ``BST problem $\to$ fragmentation problem
$\to$ MDP problem'' and use the traveling front technique to analyze the MDP
problem.  This technique allows to re-derive in a physically
transparent way all results for the RBST mentioned above and provides a
lot of new results.  We will show that the constants $\alpha_0$ and
${\alpha_0}'$ are simply related to the velocities of traveling fronts.  The
sub-leading correction terms can also be derived analytically.  The traveling
front approach also predicts `concentration of measure' of the variables
$H_N$ and $h_N$. This means that the asymptotic probability distributions of
these variables are highly localized around their respective averages. As a
result, a typical value of $H_N\sim \langle H_N\rangle$ and the spread in
$H_N$ is of order $O(1)$ in the large $N$ limit.  Naturally the variance and
higher cumulants of both $H_N$ and $h_N$ are bounded.  We also derive an
asymptotically exact nonlinear integral equation for the full probability
distributions of $H_N$ and $h_N$. While we could not solve this nonlinear
equation in closed form, we could derive the behaviors at the tails of these
highly localized distribution functions. We will also see that within this
approach the variables $H_N$ and $h_N$ map respectively onto the minimum and
maximum energy of a directed polymer and hence the observed duality between
them is rather natural.

The main advantage of the present approach is that it allows us to go beyond
the random trees and obtain exact asymptotic results for the statistics of
$H_N$ and $h_N$ for BST's with arbitrary distributions.  This is the main new
result of the present paper. Besides, we also generalize basic results to
$m$-ary search trees with arbitrary distributions.

\section{Mapping of the BST problem to a fragmentation problem}

In order to derive the asymptotics of the statistics of the height and the
balanced height in the BST problem, it is convenient to first map this
problem to a fragmentation problem following Devroye\cite{D1,D2}. To
illustrate how this mapping works, let us consider again the example in
Fig.~1 where the months (or the corresponding integers from $1$ to $12$)
appear in the particular sequence $[7,9,12,5,4,2,1,10,11,3,6,8]$. The first
element (which in this example is $7$) is chosen randomly from the available
$N=12$ elements in the case of RBST. Once this element is chosen, the
remaining elements will belong either to the interval $[1-6]$ or $[8-12]$,
which are subsequently completely disconnected from each other. Thus choosing
the first element is equivalent to breaking the original interval $[1-12]$
into two intervals, the left $[1-6]$ and the right $[8-12]$ at the break
point $7$ which is chosen randomly. Now consider the next element. It will
either belong to the left or the right interval. In the particular example we
are discussing, the next element $9$ belongs to the right interval $[8-12]$.
This new element then divides the right interval $[8-12]$ again into two
parts: the left containing only the single element $[8]$ and the right
$[10-12]$. These two new intervals subsequently become completely independent
of each other. The third element $[12]$ breaks subsequently the interval
$[10-12]$ into two parts: the left part $[10-11]$ and the right part which is
empty. Similarly the fourth element $[5]$ breaks the interval $[1-6]$ into
two parts: the left part $[1-4]$ and the right part $[6]$ and so on.

Thus one can think of the construction of the RBST as a dynamical
fragmentation process where one starts with a stick of initial length $N$ and
breaks it randomly into two parts: a left part of length $rN$ and a
right part of length $r'N$ with the constraint $r+r'=1$ where $r$ is a random
number distributed uniformly over the interval $[0-1]$. At the next step, one
breaks each of these intervals again into two parts.  At any stage of
breaking, the random variable $r$ characterizing the break point of an
interval is chosen independently from interval to interval. They are also
independent from stage to stage.  After $n$ steps of breaking, there are
$2^n$ intervals.  Note that this fragmentation process has itself a tree
structure and can be represented by a branching process as depicted in Fig.~3.

\begin{figure}
\narrowtext\centerline{\epsfxsize\columnwidth \epsfbox{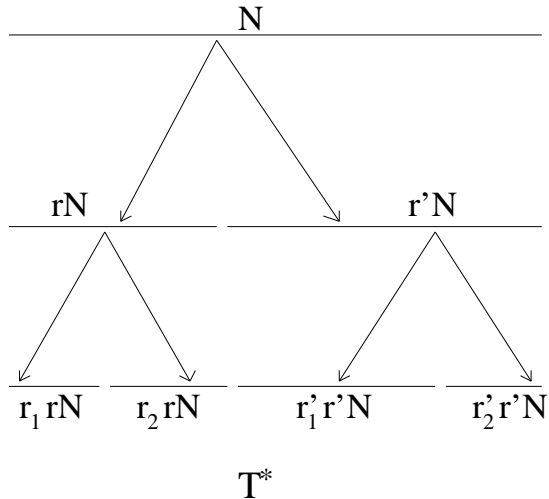}}
\caption{The fragmentation process has itself a tree structure 
  (denoted by $T^*$), shown here up to level $2$.  In the first step an
  interval of length $N$ is broken into two pieces of lengths $rN$ and $r'N$
  such that $r+r'=1$. Each of those pieces are further broken into two halves
  satisfying the constraints $r_1+r_2=1$ and ${r_1}'+{r_2}'=1$. At level $n$,
  there will be $2^n$ pieces.}
\end{figure}

A search tree of fixed size $N$ is completed when in the corresponding
fragmentation process, the lengths of all intervals are less than $1$
because this means that all the elements of the incoming data string have
already been incorporated onto the search tree.  Although in the
fragmentation problem we have continuous intervals whereas in the RBST
the intervals consist of discrete integers, it does not really matter since
one can associate the integer part of a break point to a particular integer
element of the RBST. For example, if the first break point in the
fragmentation problem is $7.3$, this means that in the RBST
problem, the first element (the root) is integer $7$.

Let us first consider the height $H_N$ of the RBST.  By definition, $H_N$ is
the distance from the root (depth) of the farthest element in the RBST. The
RBST stops growing beyond $H_N$ as all the incoming $N$ elements have been
incorporated in the tree.  Thus when the RBST attains the depth $H_N$, in the
corresponding fragmentation problem, the length of every interval is less
than $1$.  Denote by $l_1$, $\ldots$, $l_{2^n}$ the lengths of $2^n$
intervals after $n$ steps of breaking.  Clearly, the probability ${\rm
  Prob}[H_N<n]$ in the RBST problem is the same as the probability that all
$2^n$ intervals in the fragmentation problem have lengths less than $1$,
\begin{equation} 
{\rm {Prob}}\left[H_N<n\right]={\rm Prob}\left[l_1<1,\ldots,\l_{2^n}<1\right].
\label{probH}
\end{equation} 
The right hand side of Eq.~(\ref{probH}) is also the probability that the
maximum of the lengths of the $2^n$ pieces is less than $1$ in the
fragmentation problem.

We next consider the balanced height $h_N$ of the RBST. By definition, $h_N$
is the depth up to which the RBST is fully saturated and balanced. Beyond this
depth, some parts of the RBST stop growing (see Fig.~1 where $h_N=3$). This
means that in the corresponding random fragmentation process, as long as the
step number of breaking is less than $h_N$, the lengths of all the intervals
must still be bigger than $1$, so that each such interval can incorporate a
new element. Thus the probability ${\rm Prob}[h_N>n]$ in the RBST is the same
as the probability that all $2^n$ intervals in the fragmentation problem
have lengths bigger than $1$, 
\begin{equation} 
{\rm Prob}\left[h_N>n\right]={\rm
  Prob}\left[l_1>1, \ldots,\l_{2^n}>1\right].
\label{probh}
\end{equation} 
The right hand side of Eq.~(\ref{probh}) is also the probability that the
minimum of the lengths of the $2^n$ pieces is bigger than $1$ in the
fragmentation problem.

In the RBST, the new elements in the tree arrive randomly. The corresponding
fragmentation problem is also random in the sense that at each stage an
interval $l$ is broken into two parts of lengths $rl$ and $r'l$ with $r+r'=1$
where the random variable $r$ is chosen each time independently and is
distributed uniformly over $[0-1]$. One can, of course, generalize this
random fragmentation problem where the variable $r$ is chosen independently
each time but with an arbitrary distribution over $[0-1]$, not necessarily
uniform. This would correspond to a BST problem where the new elements arrive
with a specified distribution. In general, at any stage of breaking, the
joint probability distribution of $r$ and $r'$ can be written as
\begin{equation} 
{\rm Prob}\left[r,r'\right]= \phi(r) \phi(r') \delta(r+r'-1).
\label{jointr}
\end{equation} 
The delta function ensures that the total length is conserved at every stage
of breaking.  The joint distribution is written in a symmetric way to ensure
that both $r$ and $r'$ have the same effective distribution which is given by
$\eta(r)={\rm Prob}(r)=\int_{0}^{1}{\rm Prob}[r,r']dr'= \phi(r)\phi(1-r)$.
The function $\phi(r)$ must be chosen such that the induced distribution
$\eta(r)$ satisfies the conditions, $\int_0^{1}\eta(r)dr=1$ and $\int_0^{1}
r\eta(r)dr=1/2$. The first condition condition ensures normalizability of the
single point distribution $\eta(r)$ and the second condition comes from the
strict constraint $r+r'=1$ which indicates $\langle r\rangle=\langle
r'\rangle=1/2$. In the case of random breaking, the function $\phi(r)=1$ and
consequently the induced distribution $\eta(r)=1$ for $0\le r\le 1$.  A
simple example of a non-random break point distribution is given by,
$\phi(r)=\sqrt{6}\,r$ with the induced distribution $\eta(r)=6r(1-r)$ that
satisfies the two constraints\cite{KM}.

Apart from connection to the BST problem, the random fragmentation problem is
interesting on its own rights as it arises in various contexts such
as the energy cascades in turbulence\cite{GEL}, rapture processes in
earthquakes\cite{NG}, financial crashes in stock markets\cite{SJ}, and in the
stress propagation in granular medium\cite{qmodel}. In our previous
paper\cite{KM}, we had studied the asymptotic laws governing the probability
distribution of the maximal lengths of the intervals after $n$ steps of
breaking in the random fragmentation problem using traveling front
techniques.  The same differential equation that describes the Laplace
transform of this distribution was also studied independently by Drmota via a
different method\cite{Drmota}.  Both these methods work well for the random
problem (where $\eta(r)=1$) but seem difficult to extend to the general case
when the break point in the fragmentation process is chosen from an arbitrary
induced distribution $\eta(r)$\cite{KM}. It turns out, however, that the
fragmentation problem with general $\eta(r)$ can further be mapped to a MDP
problem as presented in the next section. This further mapping followed by
the traveling front analysis then allows us to obtain exact asymptotic
results for the general case with arbitrary $\eta(r)$.
 
\section{Mapping of the fragmentation problem to a modified 
directed polymer problem}

In this section we further map the fragmentation problem onto a MDP problem
on a Cayley tree. This MDP problem turns out to be slightly different from
the conventional DP problem studied in statistical mechanics due to the
presence of a special constraint. Nevertheless, asymptotic properties in the
MDP problem can be derived analytically using the traveling front techniques.

To understand this mapping, consider the set of $2^n$ intervals in the
fragmentation problem after $n$ steps of breaking, starting from the initial
length $N$. Let $l_k$ denote the length of the $k$-th interval where $k=1$,
$\ldots$, $2^n$. From Fig.~3, it is clear that the length of any typical
piece $l_k$ can be expressed as the product 
\begin{equation}
l_k= N \prod_{i=1}^{n} r_i \,
,
\label{prodl}
\end{equation} 
where $r_i$'s are the set of independent random variables encountered in
getting the final piece of length $l_k$ after $n$ steps of breaking the
original interval of length $N$. Note that in the tree $T^{*}$ in Fig.~3,
there is a unique path connecting the original interval (the root element of
$T^{*}$) to the $k$-th interval at stage $n$ and the set of random variables
$r_i$'s encountered in going from the root of $T^{*}$ to the $k$-th piece at
stage $n$ defines this unique path.  Alternately we can associate an energy
variable $\epsilon_i=-\log r_i \ge 0$ to the bonds connecting this path and
the set of energies $\epsilon_i$'s also uniquely characterize the path (see
Fig.~4). Taking logarithm in Eq.~(\ref{prodl}), we see that the total energy
$E_k$ of a path (starting at the root and ending at the $k$-interval at the
stage $n$) becomes
\begin{equation} 
E_k = \log \left({{N}\over
    {l_k}}\right)=-\sum_{i=1}^{n} \log r_i = \sum_{i=1}^n \epsilon_i .
\label{Ek}
\end{equation} 
This path then represents a typical configuration of a directed polymer
(directed in the downward direction) with energy given by Eq.~(\ref{Ek})
where $\epsilon_i$'s are random bond energies.  Note that up to level $n$,
there are a total number of $2^n$ different paths each having different total
energies $E_1$, $\ldots$, $E_{2^n}$.

In the conventional DP problem, the bond energies $\epsilon_i$'s are
completely uncorrelated. To understand why they are correlated in the present
problem, recall that when an interval is broken into two parts the random
variables $r$ and $r'$ characterizing the lengths of the two daughter
intervals satisfy the length conservation constraint, $r+r'=1$.  Translated
into the DP problem, the corresponding bond energies $\epsilon=-\log r$ and
$\epsilon'=-\log r'$ associated with the two bonds emanating downwards from a
given node must satisfy the constraint
\begin{equation} 
e^{-\epsilon} +
e^{-\epsilon'}=1.
\label{constraint}
\end{equation} 
This constraint holds at every branching point of the tree (see Fig.~4).
This correlation makes the MDP problem slightly different from the
conventional DP problem.  

The joint distribution $p(\epsilon,\epsilon')$
of the energies of the two bonds emanating from the common node and  
the induced effective single bond distribution $\rho(\epsilon)$ are obtained
from Eq.~(\ref{jointr}) to give:
\begin{eqnarray}
p(\epsilon,\epsilon')
&=&\phi\left(e^{-\epsilon}\right)\phi\left(e^{-\epsilon'}\right)
e^{-\epsilon-\epsilon'}
\delta\left(e^{-\epsilon}+e^{-\epsilon'}-1\right),\nonumber \\
\rho(\epsilon)&\equiv &\int_0^{\infty} p(\epsilon,\epsilon')d\epsilon' 
=\phi\left(e^{-\epsilon}\right)\phi\left(1-e^{-\epsilon}\right)e^{-\epsilon}.
\label{rhoe}
\end{eqnarray} 
For example, for the RBST we have $\phi(r)=1$, and therefore
$\rho(\epsilon)=e^{-\epsilon}$.  Note that in the conventional DP problem,
the joint distribution $p(\epsilon,\epsilon')$ would simply be the product of
the single point distributions,
$p(\epsilon,\epsilon')=\rho(\epsilon)\rho(\epsilon')$ since they are
independent. The MDP problem, however, lacks this factorization property.

\begin{figure}
  \narrowtext\centerline{\epsfxsize\columnwidth \epsfbox{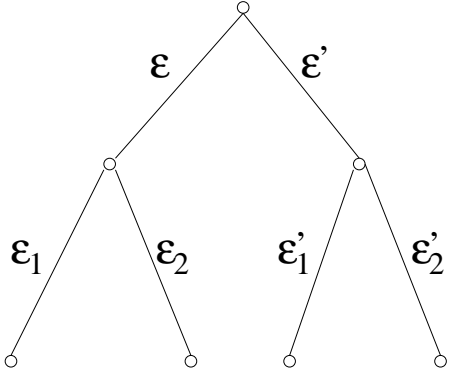}}
\caption{The MDP on a Cayley tree. This tree is isomorphic to the
  tree of the fragmentation process shown in Fig.~3. Each bond energy
  $\epsilon$ is related to the corresponding fraction $r$ via $\epsilon=-\log
  r$. The bond energies are correlated due to the constraints:
  $e^{-\epsilon}+e^{-\epsilon'}=1$, $e^{-\epsilon_1}+e^{-\epsilon_2}=1$,
  $e^{-{\epsilon_1}'} +e^{-{\epsilon_2}'}=1$, etc.}
\end{figure}
\noindent  

Having set up the notation we turn to the variables $H_N$ and $h_N$ in
the original BST problem. What do the distributions of $H_N$ and $h_N$
correspond to in the MDP problem?  First, consider the height
distribution ${\rm Prob}\left[H_N<n\right]$.  {}From Eqs.~(\ref{probH}) and
(\ref{Ek}), one finds
\begin{eqnarray} 
{\rm {Prob}}\left[H_N<n\right]={\rm Prob}\left[l_1<1,
  \ldots,\l_{2^n}<1\right] \nonumber \\
= {\rm Prob}\left[E_1>\log N, \ldots,E_{2^n}>\log N\right],
\label{probH2}
\end{eqnarray} 
where $E_k$'s ($k=1$, $\ldots$, $2^n$) are respectively the total
energies of the all possible $2^n$ paths going from the root to the leaves at
the $n$-th level in the DP problem.  The probability in the last line in Eq.
(\ref{probH2}) is also the same as the probability ${\rm Prob}\left[{\rm
    min}\{E_1, \ldots, E_{2^n}\}>\log N \right]$. Thus the height
distribution ${\rm Prob}\left[H_N<n\right]$ in the BST problem is precisely
related to the distribution of the minimum (ground state) energy of the MDP
problem, a quantity of considerable interest in statistical physics.

Let us next consider the balanced height $h_N$. Using Eqs.~(\ref{probh}) and
(\ref{Ek}), it follows similarly that, 
\begin{eqnarray} 
{\rm {Prob}}\left[h_N>n\right]={\rm Prob}\left[l_1>1,
  \ldots,\l_{2^n}>1\right] \nonumber \\
= {\rm Prob}\left[E_1<\log N, \ldots,\, E_{2^n}<\log N\right],
\label{probh2}
\end{eqnarray} 
which is also the probability that the maximum energy ${\rm max}\{E_1,
\ldots, E_{2^n}\}$ is less than $\log N$.  Thus the balanced height
distribution ${\rm Prob}\left[h_N>n\right]$ in the BST problem is related to
the distribution of the maximum energy in the MDP problem, a quantity that is
usually not of much interest in statistical mechanics.

\subsection{Statistics of the Height or the Minimum Energy}

In this subsection we analyze the asymptotic statistics of the height $H_N$
in the BST problem or equivalently that of the minimum energy in the MDP
problem.  Let $P_n(x)={\rm Prob}\left[{\rm min}\{E_1, \ldots, E_{2^n}\}>x
\right]$, where $E_k$'s with $k=1$, $\ldots$, $2^n$ are the energies of the
$2^n$ polymer paths from the root to the $n$-th level. It is then easy to
write a recursion relation for $P_n(x)$,
\begin{equation}
P_{n+1}(x)=\int_0^{\infty}\int_0^{\infty}
P_{n}(x-\epsilon)P_{n}(x-\epsilon')\,p(\epsilon,\epsilon')\,d\epsilon
\,d\epsilon' \, ,
\label{recur1}
\end{equation} 
where $p(\epsilon,\epsilon')$ is the joint distribution of the two bond
energies as given by Eq.~(\ref{rhoe}).  Equation (\ref{recur1}) has been
derived by analyzing different possibilities for the energies of the bonds
emanating from the root and using the fact that the two subsequent daughter
trees are statistically independent. Note that in the conventional DP
problem, the corresponding recursion relation would be simplified using the
factorization property of the joint distribution $p(\epsilon,\epsilon')$ and
one would get\cite{MK,DM},
\begin{equation}
P_{n+1}(x)=\left[\int_0^{\infty}
  P_{n}(x-\epsilon)\,\rho(\epsilon)\,d\epsilon\right]^2 .
\label{recurcdp}
\end{equation}

We have to solve the recursion relation (\ref{recur1}) subject to the initial
condition
\begin{equation}
P_0(x)=\cases{1 &$x\leq 0$,\cr 
              0 &$x>0$,\cr}
\label{P0}
\end{equation} 
and the boundary conditions 
\begin{equation}
P_n(x)\to\cases{1 &  $x\to -\infty$,\cr 
                0 &  $x\to \infty$.\cr}
\label{Pbound}
\end{equation} 
The recursion relation (\ref{recur1}) is nonlinear and in general difficult
to solve exactly.  However, its asymptotic properties can be derived
analytically. As $n$ increases, the solution $P_n(x)$ in Eq.~(\ref{recur1})
looks like a $[1-0]$ front (i.e., $P_n(x)\sim 1$ for small x but falls off
rapidly to $0$ for large $x$) advancing in the positive direction. This
suggests that for large $n$, Eq.~(\ref{recur1}) admits a traveling front
solution, $P_n(x)=F(x-x_n)$ where $x_n$ denotes the location of the front and
the shape of the front is described by the fixed point scaling function $F$
that becomes independent of $n$. This implies that the width of the front is
of order $O(1)$, i.e., it saturates in the large $n$ limit.  The traveling
front ansatz also indicates that the front advances with a uniform velocity,
i.e. $x_n\approx vn$, to leading order for large $n$ where the velocity $v$
is yet to be determined. Substituting this traveling front ansatz,
$P_n(x)=F(x-vn)$ for large $n$ in Eq.~(\ref{recur1}), we find that the fixed
point function $F(y)$ satisfies the nonlinear integral equation,
\begin{equation} 
F(y-v)=\int_0^{\infty}\int_0^{\infty}F(y-\epsilon)\,
F(y-\epsilon')\,p(\epsilon,\epsilon')\,d\epsilon\, d\epsilon' \, ,
\label{Fy}
\end{equation} 
where the velocity $v$ is still undetermined and $F(y)$ satisfies the
boundary conditions 
\begin{equation}
F(y)\to\cases{1 & as $y\to -\infty$,\cr 
              0 &as $y\to \infty$.\cr}
\label{Fbound}
\end{equation} 

Let us first analyze Eq.~(\ref{Fy}) in the tail region $y\to -\infty$.
Plugging $F(y)=1-f(y)$ in Eq.~(\ref{Fy}) and neglecting terms of order
$O(f^2)$ we find that $f(y)$ satisfies
\begin{equation} 
f(y-v)=2\int_0^{\infty}
f(y-\epsilon)\,\rho(\epsilon)\,d\epsilon,
\label{fy}
\end{equation} 
where we have used the relation $\rho(\epsilon)=\int_0^{\infty}
p(\epsilon, \epsilon')d\epsilon'$. This linear equation (\ref{fy}) clearly
admits an exponential solution $f(y)=\exp(\lambda y)$ provided the inverse
decay rate $\lambda$ is related to the velocity $v$ via the dispersion
relation 
\begin{equation} 
v(\lambda)= -{1\over {\lambda}}\log \left[
  2\int_0^{\infty}e^{-\lambda \epsilon}\,\rho(\epsilon)\,d\epsilon\right].
\label{vl}
\end{equation} 
For a given induced distribution $\rho(\epsilon)$, the function
$v(\lambda)\to -\log(2)/{\lambda}$ as $\lambda\to 0$ and $v(\lambda)\to 0$ as
$\lambda\to \infty$ with a single maximum at a finite $\lambda^{*}$
determined via 
\begin{equation} 
{{dv}\over {d\lambda}}\Big|_{\lambda^{*}}=0.
\label{l*}
\end{equation}

Thus for all $\lambda$ such that $\int_0^{\infty}
e^{-\lambda\epsilon}\rho(\epsilon)\,d\epsilon<1/2$, the corresponding
velocity $v(\lambda)>0$. While any such $\lambda$ is in principle allowed, a
particular velocity is actually asymptotically selected by the front. This
velocity selection mechanism has been observed in a large class of nonlinear
problems with a traveling front solution
\cite{Fisher,KPP,Bramson,Zel,Murray,VS1,VS2,KM,BD,MK,DM}. It is known that as
long as the initial condition is sharp (as in the present case in
Eq.~(\ref{P0})), the extreme value is chosen.  {}From this general front
selection principle, we infer that in our present problem, the front finally
selects the velocity $v(\lambda^*)$ where $\lambda^*$ is given by the
solution of Eq.~(\ref{l*}).  Thus the asymptotic front position, to
leading order for large $n$, is given by
\begin{equation} 
x_n \approx v(\lambda^*)\,n.
\label{xn}
\end{equation} 

While the leading behavior of the front position $x_n$ is given exactly by
Eq.~(\ref{xn}), it turns out that it has an associated slow logarithmic
correction. This logarithmic correction to the front velocity was first
derived by Bramson in the context of a reaction-diffusion equation
\cite{Bramson}, and was subsequently found in many other systems with a
traveling front\cite{VS1,VS2,KM,BD,MK}.  In Appendix A, we present a
detailed derivation of this correction term following the approach of Brunet
and Derrida\cite{BD}.  The main result of this exercise is that the
asymptotic front position for large $n$ is given by
\begin{equation}
x_n \approx
v(\lambda^*)\,n + {3\over {2\lambda^*}}\,\log n .
\label{xn2}
\end{equation} 
One can even calculate the next correction term by employing a more
sophisticated approach\cite{VS2} but we omit these results here. One
important point to note is that while the velocity $v(\lambda^*)$ and
$\lambda^*$ are nonuniversal as they depend explicitly on the distribution
$\rho(\epsilon)$, the prefactor $3/2$ of the logarithmic correction in
Eq.~(\ref{xn2}) is actually universal and is precisely the first excited
state energy of a quantum harmonic oscillator (see Appendix A).

Let us now translate back these results to see what they mean for the height
distribution in the original BST problem. {} From Eq.~(\ref{probH2}), it is
clear that the cumulative height distribution for large $n$ is given by 
\begin{equation}
{\rm Prob}\left[H_N<n\right]=P_n(\log N)\approx F(\log N - x_n),
\label{cdistH}
\end{equation} 
where the front position $x_n$ is given by Eq.~(\ref{xn2}) and the
function $F(y)$ is given by the solution of Eq.~(\ref{Fy}). Since the
function $F(y)$ has the shape of a front with center at $y=0$ and width of
order $O(1)$, its derivative $F'(y)$ is a localized function around $y=0$
with width of order $O(1)$. From Eq.~(\ref{cdistH}) it then follows that the
height distribution ${\rm Prob}\left[H_N=n\right]$ is also localized around
its average value $\langle H_N\rangle$ with a variance $V(H_N)\sim O(1)$.
Thus $H_N$ has a concentration of measure around its average value $\langle
H_N\rangle$ which is given by the value of $n$ that corresponds to the zero
of the argument of the function $F(y)$, i.e., when $x_n=\log N$.  Using
$x_n=\log N$ in Eq.~(\ref{xn2}) and solving for the required value of $n$ for
large $N$, we obtain one of our main results 
\begin{equation} 
\langle H_N\rangle = {1\over {v(\lambda^*)}}\,\log N 
- {3\over {2\lambda^*v(\lambda^*)}}\,\log \left(\log N
\right),
\label{H_N}
\end{equation} 
where $v(\lambda)$ and $\lambda^*$ are given respectively by Eqs.
(\ref{vl}) and (\ref{l*}). This is the first result for the fragmentation
problem with arbitrary break-point distribution going beyond the uniform case
or equivalently for the BST problem where the elements in the tree arrive
with an arbitrary distribution and not just randomly.

It is useful to exemplify the above general results. For the original RBST
problem, $\phi(r)=1$ or $\rho(\epsilon)=e^{-\epsilon}$ [see
Eq.~(\ref{rhoe})].  Substituting $\rho(\epsilon)=e^{-\epsilon}$ into
Eq.~(\ref{vl}) we get
\begin{equation}  
v(\lambda)= -{1\over {\lambda}}\log \left[ {2\over {\lambda+1}}\right],
\label{vlr}
\end{equation} 
which has a single maximum at $\lambda^*=3.31107\ldots$ with
$v(\lambda^*)=0.23196\ldots$. Substituting $\lambda^*$ and $v(\lambda^*)$
into Eq.~(\ref{H_N}) we arrive at Eq.~(\ref{HN1}) with
$\alpha_0=4.31107\ldots$ and $\alpha_1=-1.95302\ldots$, in agreement with
Refs\cite{KM,Reed}.
 
Consider another example, $\phi(r)={\sqrt 6}\,r$, a problem that couldn't be
solved by the techniques used in our previous short paper\cite{KM}.  This
corresponds to the fragmentation problem where the induced distribution of
the break-point is $\eta(r)=6r(1-r)$. In the MDP problem, it corresponds to
the induced bond energy distribution
\begin{eqnarray} 
\label{rhoe1} 
\rho(\epsilon)=6\,e^{-2\epsilon}(1-e^{-\epsilon}).  
\end{eqnarray} 
Substituting this form
in Eq.~(\ref{vl}), we get
\begin{eqnarray} 
\label{vl2} 
v(\lambda)=-{1\over
  {\lambda}}\log \left[ {12\over {(\lambda+2)(\lambda+3)}}\right],
\end{eqnarray} 
which has a unique maximum at $\lambda^*=3.92408\ldots$ and this maximum
velocity is given by $v(\lambda^*)=0.31322\ldots$. Substituting these results
into the general formula (\ref{H_N}) we recover Eq.~(\ref{HN1}) with 
$\alpha_0=3.19258\dots$ and $\alpha_1=-1.22038\dots$. 

The traveling front approach also gives the full probability distribution of
the height variable in the BST and not just its exact average value as in
Eq.~(\ref{H_N}). Indeed, we have seen that cumulative height distribution is
given by Eq.~(\ref{cdistH}) where the function $F(y)$ is the solution of the
boundary value problem (\ref{Fy})--(\ref{Fbound}).  While we have not been
able to solve the nonlinear integral equation (\ref{Fy}) exactly, one can
easily deduce the extreme behavior of $F(y)$.  We have already seen that in
the tail region $y\to -\infty$, the function $F(y)$ saturates to $1$
exponentially fast, $1-F(y)\sim \exp[\lambda^*y]$, where $\lambda^*$ is the
solution of Eq.~(\ref{l*}). One can also deduce the asymptotic behavior of
$F(y)$ when $y\to \infty$ (Appendix B) for arbitrary distribution
$\rho(\epsilon)$.  Thus the asymptotic behaviors of the function $F(y)$ read
\begin{equation}  
F(y)\approx \cases{1- A e^{\lambda^* y} &$y\to -\infty$,\cr 
2\int_y^{\infty}\rho(y'+v)dy' &$y\to \infty$,\cr}
\label{Fy5}
\end{equation} 
where $A$ is a constant, $\lambda^*$ is found from
(\ref{l*}), and $v=v(\lambda^*)$. In particular, for the RBST where
$\rho(\epsilon)=e^{-\epsilon}$, $\lambda^*=3.31107$, and
$v(\lambda^*)=0.23196$, one has 
\begin{equation}  
F(y)\approx \cases{ 1- A e^{3.31107 y} &$y\to
  -\infty$,\cr 1.58596 e^{-y} &$y\to \infty$.\cr}
\label{Fy6}
\end{equation}

In conclusion, the height distribution is a localized function around its
average value $\langle H_N\rangle$ given by Eq.~(\ref{H_N}).  For any
unbounded distribution $\rho(\epsilon)$, the height distribution decays at in
the tail regions according to Eq.~(\ref{Fy5}). For bounded distributions,
however, $F(y)$ vanishes for sufficiently large $y$.  Recall that the
distribution of the minimum of a set of uncorrelated random variables is
known to have a universal superexponential decay for large
value\cite{Gumbel}.  However it was shown in Ref. \cite{DM} that in the
conventional DP problem the distribution of the minimum energy of a polymer
violates this Gumbel law due to hierarchical correlations between the
energies of different paths.  From Eq.~(\ref{Fy5}) it is clear that in the
MDP problem the forward tail is nonuniversal since it depends explicitly on
the distribution $\rho(\epsilon)$.  Generally, the forward tail is not
superexponential thus clearly violating the Gumbel statistics.
  
\subsection{Statistics of the Balanced Height or the Maximum Energy}

The analysis of the statistics of balanced height $\langle h_N\rangle$
follows more or less the same approach as in the case of height variable,
except that one is now concerned with the distribution of maximum energy in
the MDP problem. Let $R_n(x)={\rm Prob}\left[{\rm max}\{E_1, E_2, \ldots,
  E_{2^n}\}<x \right]$ where $E_k$'s with $k=1$, $2$, $\ldots$, $2^n$ are the
energies of the $2^n$ polymer paths from the root to the $n$-th level. Then
$R_n(x)$ satisfies the same recursion relation as the $P_n(x)$ in 
Eq.~(\ref{recur1}), 
\begin{equation}
R_{n+1}(x)=\int_0^{\infty}\int_0^{\infty}R_n(x-\epsilon)R_n(x-\epsilon')
p(\epsilon,\epsilon')d\epsilon d\epsilon' \, .
\label{recur2}
\end{equation} 
The only difference is in the initial condition, 
\begin{equation} 
R_0(x)=\cases{0 &$x\leq 0$,\cr 
              1 &$x>0$,\cr}
\label{R0}
\end{equation} 
and in the boundary conditions, 
\begin{equation}
R_n(x)\to\cases{0 &  $x\to -\infty$,\cr 
                1 &  $x\to \infty$.\cr}
\label{Rbound}
\end{equation} 
As in the case of Eq.~(\ref{recur1}), the recursion relation (\ref{recur2})
admits a traveling front solution for large $n$, $R_n(x)=G(x-{x_n}^*)$ where
${x_n}^*$ is the front position and the fixed point scaling function function
$G(x)$ describes the shape of the front.  Unlike the $[1-0]$ front in the
previous subsection, the front for $R_n(x)$ has a $[0-1]$ form advancing in
the positive direction.  The front again advances with asymptotically
constant velocity $v_1$, i.e., the position of the front is ${x_n}^*\approx
v_1 n$. Substituting $R_n(x)=G(x-v_1n)$ in Eq.~(\ref{recur2}), we find that
$G(y)$ satisfies the nonlinear integral equation
\begin{equation}
G(y-v_1)=\int_0^{\infty}\int_0^{\infty}G(y-\epsilon)
G(y-\epsilon')p(\epsilon,\epsilon')d\epsilon\, d\epsilon' \,.
\label{Gy}
\end{equation} 
The velocity $v_1$ is still undetermined and the front shape $G(y)$ satisfies
the boundary conditions, $G(y)\to 0$ as $y\to -\infty$ and $G(y)\to 1$ for
$y\to \infty$.  As in the previous subsection, we will analyze the
Eq.~(\ref{Gy}) in the tail where $G(y)\to 1$; in the present case, this means
$y\to \infty$.  Substituting $G(y)=1-g(y)$ in Eq.~(\ref{Gy}) and neglecting
terms of order $O(g^2)$ we get the linear equation
\begin{equation}
g(y-v_1)=2\int_0^{\infty} g(y-\epsilon)\,\rho(\epsilon)\,d\epsilon.
\label{gy}
\end{equation} 
Equation (\ref{gy}) admits asymptotically exponential solution,
$g(y)=\exp(-\mu y)$ as $y\to \infty$, with
\begin{equation} 
v_1(\mu)= {1\over {\mu}}\,\log \left[ 2\int_0^{\infty}e^{\mu
    \epsilon}\rho(\epsilon)d\epsilon\right].
\label{vm}
\end{equation} 
The dispersion relation in Eq.~(\ref{vm}) has a single minimum at
$\mu=\mu^*$ determined from relation
\begin{equation}
{{dv_1}\over {d\mu}}\Big|_{\mu^{*}}=0.
\label{m*}
\end{equation} 
By the general front selection mechanism, we infer that this
minimum velocity will be selected by the front 
\begin{equation} 
{x_n}^* \approx
v_1(\mu^*)n .
\label{xn*}
\end{equation} 
The associated slow logarithmic correction can also be worked out
following the same calculation as in Appendix A and we finally get 
\begin{equation}
{x_n}^* \approx v_1(\mu^*)\,n - {3\over {2\mu^*}}\,\log n .
\label{xn*2}
\end{equation} 
Note that the correction term in Eq.~(\ref{xn*2}) has a negative sign
compared to the positive sign in Eq.~(\ref{xn2}).

In terms of the BST problem, it is clear from Eq.~(\ref{probh2}) that the
cumulative balanced height distribution for large $n$ is given by 
\begin{equation}
{\rm Prob}\left[h_N>n\right]=R_n(\log N)\approx G(\log N - {x_n}^*),
\label{cdisth}
\end{equation} 
where the front position ${x_n}^*$ is given by Eq.~(\ref{xn*2}) and the
function $G(y)$ is the solution of Eq.~(\ref{Gy}). As argued in the previous
subsection, the derivative $G'(y)$ is a localized function
around $y=0$ with width of order $O(1)$.  Thus
the balanced height distribution ${\rm Prob}\left[h_N=n\right]$ is also
localized around its average value $\langle h_N\rangle$ with a variance
$V(h_N)\sim O(1)$.  The average value reads
\begin{equation} 
\langle h_N\rangle = {1\over {v_1(\mu^*)}}\,\log N
+ {3\over {2\mu^*v_1(\mu^*)}}\,\log \left(\log N \right).
\label{h_N}
\end{equation} 

Consider again the same examples that were studied for the
height variable in the previous subsection. For the RBST problem where
$\phi(r)=1$ or equivalently $\rho(\epsilon)=e^{-\epsilon}$, 
Eq.~(\ref{vm}) becomes 
\begin{equation}
v_1(\mu)= {1\over {\mu}}\,\log \left[ {2\over {1-\mu}}\right],
\label{vmr}
\end{equation} 
which has a single minimum at $\mu^*=0.62663\ldots$ where
$v_1(\mu^*)=2.67834\ldots$.  Thus, Eq.~(\ref{h_N}) reduces to Eq.~(\ref{hN1}) 
with ${\alpha_0}'=0.37336\ldots$ and
${\alpha_1}'=0.89374\ldots$.

For the second example, $\phi(r)={\sqrt 6}\,r$ or equivalently
$\rho(\epsilon)$ given by Eq.~(\ref{rhoe1}), Eq.~(\ref{vm}) becomes 
\begin{equation}
v_1(\mu)={1\over {\mu}}\log \left[ {12\over {(2-\mu)(3-\mu)}}\right].
\label{vm2}
\end{equation} 
The function $v_1(\mu)$ in Eq.~(\ref{vm2}) has a single minimum at
$\mu^*=1.17864\ldots$ where $v_1(\mu^*)=1.76653\ldots$.  Equation (\ref{h_N})
again reduces to Eq.~(\ref{hN1}) with ${\alpha_0}'=0.56607\ldots$ and
${\alpha_1}'=0.72041\ldots$.

Finally we explain the duality between $H_N$ and $h_N$ in the BST problem. In
the language of the MDP problem, these variables correspond to the minimum
and maximum energy of a directed polymer in a random medium where the bond
energies $\epsilon_i$'s have nonzero support only for $\epsilon_i\ge 0$.
Changing the sign of the bond energies maps the minimum energy in the
negative support problem into the negative of the maximum energy in the
positive support problem.  This fact is reflected in the relation between the
two dispersion relations in Eqs.~(\ref{vl}) and (\ref{vm}),
$v(-\lambda)=v_1(\lambda)$.  Thus $\lambda^*$ and $-\mu^*$ are actually the
two different roots of the same transcendental equation (\ref{l*}).
Consequently, the constants $\alpha_0$ and ${\alpha_0}'$ in
Eqs.~(\ref{HN1})--(\ref{hN1}) are merely two different roots of the same
transcendental equation.

\section{Generalization to $m$-ary search trees with arbitrary distributions}

The results obtained in the previous sections for the statistics of $H_N$ and
$h_N$ of the BST's with arbitrary distributions can be generalized in a
straightforward manner to the $m$-ary search trees. An $m$-ary search tree is
constructed in the following way.  One first collects the first $(m-1)$
elements of the incoming data string and arranges them together in the root
of the tree in an ordered sequence $x_1<\ldots<x_{m-1}$.  Next when the
$m$-th element $x_m$ comes, one compares first with $x_1$. If $x_m<x_1$, the
$m$-th element is assigned to the root of the leftmost daughter tree. If
$x_1<x_m<x_2$, then $x_m$ goes to form the root of the second branch and so
on. Each subsequent incoming element is assigned to either of the $m$
branches according to the above rule. Note that the level of the tree will
increase beyond a given node only when the node gets filled beyond its
capacity of $(m-1)$ elements.  Thus in the $m$-ary search tree, each node
will contain at the most $(m-1)$ elements.

The mapping to the fragmentation problem goes through following the same line
of arguments used for the binary tree in Sect.~III. In this case, one
starts with an interval of size $N$ and breaks it into $m$ pieces.
Subsequently each piece is further broken into $m$ pieces and so on.  When an
interval is broken into $m$ pieces, each of the new pieces is a fraction of
the original piece. The lengths of these $m$ new pieces are characterized by
a set of $m$ random numbers $\{r_1,\ldots,r_m\}$ such that $\sum_{i=1}^m
r_i=1$ thus enforcing the length conservation. For each interval a new set of
$r_i$'s are chosen from the same joint probability distribution 
\begin{equation} 
{\rm Prob}[r_1,\ldots,r_m]=
\delta\left(\sum_{i=1}^m r_i-1\right) \prod_{j=1}^m\phi(r_j).
\label{jointrm}
\end{equation} 
As in the binary case, the distribution (\ref{jointrm}) is written in a
symmetric form.  Note that each new piece has the same effective induced
distribution $\eta(r)$ given by the integral
$\int_0^{1}dr_2\ldots\int_0^{1}dr_m {\rm Prob}[r,r_2,\ldots,r_m]$, or
\begin{eqnarray*}
\eta(r)=\phi(r)\int_0^{1}\ldots\int_0^1 \delta\left(\sum_{i=2}^m
  r_i+r-1\right)\prod_{i=2}^{m}\phi(r_i)dr_i.
\end{eqnarray*} 
The function $\phi(r)$ must be chosen such that $\eta(r)$ satisfies the
conditions, $\int_0^1 \eta(r)dr=1$ and $\int_0^{1}r\eta(r)dr=1/m$.

The random $m$-ary search tree corresponds to a random fragmentation problem
where each of the fractions $r_1$, $\ldots$, $r_{m-1}$ is chosen from
a uniform distribution between $0$ and $1$, setting
$r_m=1-\sum_{i=1}^{m-1}r_i$, and then keeping only those sets where $r_m\ge
0$. This is precisely the so-called `uniform' distribution used by
Coppersmith et. al. in the context of the $q$-model of force fluctuations in
granular media\cite{qmodel}. In this case, $\phi(r)$ is a constant chosen
in such a way that the joint distribution (\ref{jointrm}) is
normalized.  One finds\cite{qmodel} 
\begin{equation}
{\rm Prob}[r_1,\ldots,r_m]=(m-1)!\delta\left(\sum_{i=1}^m r_i-1\right).
\label{jointrm1}
\end{equation} 
The corresponding effective single point distribution $\eta(r)$
reads\cite{qmodel}
\begin{equation}
\eta(r)=(m-1)(1-r)^{m-2}.
\label{etar1}
\end{equation}

Another interesting distribution is $\phi(r)\propto r$. In this case, the
normalized joint distribution is given by (see Appendix C)
\begin{equation} 
{\rm Prob}[r_1,\ldots,r_m]=\Gamma(2m)\,\delta\left(\sum_{i=1}^m
  r_i-1\right)\prod_{i=1}^m r_i.
\label{jointrm2}
\end{equation} 
The corresponding effective distribution $\eta(r)$ can be deduced by
recursively method (as shown in Appendix C) and we get
\begin{equation}
\eta(r)=(2m-1)(2m-2)r(1-r)^{2m-3}.
\label{etar2}
\end{equation} 
Note that for $m=2$, it reduces to $\eta(r)=6r(1-r)$ which was studied in
detail for the binary case in section III.

The $m$-piece fragmentation problem for the special case of uniform
distribution (\ref{jointrm1}) was studied in Ref. \cite{KM}. However, as in
the binary case, this method is not easy to extend to handle the general
distribution $\eta(r)$ including, for example, the distribution
(\ref{etar2}).  To go beyond the uniform case, we first map the fragmentation
problem into the MDP problem as in the binary case. One proceeds exactly as
in the binary case by associating an energy $\epsilon_i=-\log r_i$ to each
bonds of a directed polymer going from the root to the leaves of a Cayley
tree, but now with $m$ daughters emerging from each node. The energies of the
$m$ bonds emanating downwards from any given node are correlated due to the
relation $\sum_{i=1}^m r_i=1$ which translates into the constraint
\begin{equation}  
\sum_{i=1}^n e^{-\epsilon_i}=1.
\label{constraint1}
\end{equation} 
As in the binary case, this constraint holds at every branching point of the
tree. The joint distribution $p(\epsilon_1,\ldots,\epsilon_m)$ is
found from Eq.~(\ref{jointrm}) to give
\begin{equation} 
p(\epsilon_1,\ldots,\epsilon_m)=
\delta\left(\sum_{i=1}^m e^{-\epsilon_i}-1\right) \prod_{i=1}^m
e^{-\epsilon_i} \phi_i\left(e^{-\epsilon_i}\right).
\label{jointem}
\end{equation} 
Also the induced bond energy distribution $\rho(\epsilon)$ is related to
the induced fraction distribution $\eta(r)$ via 
\begin{eqnarray}
\rho(\epsilon)&=&\int_0^{\infty}\ldots \int_0^{\infty}
p(\epsilon,\epsilon_2,\ldots,\epsilon_m)
d\epsilon_2\ldots d\epsilon_m \nonumber \\
&=& \eta\left(e^{-\epsilon}\right)e^{-\epsilon}.
\label{rhoem}
\end{eqnarray}

On this $m$-branch Cayley tree, there are a total of $m^n$ possible paths of
the directed polymer going from the root to the leaves at the $n$-th level.
Following arguments similar to the binary case, the cumulative height
distribution in the $m$-ary search tree is related exactly to the
distribution of the minimum energy of the $m^n$ polymer paths in the MDP
problem via 
\begin{eqnarray} 
{\rm {Prob}}\left[H_N<n\right]={\rm Prob}\left[l_1<1,
  \ldots,\l_{m^n}<1\right] \nonumber \\
= {\rm Prob}\left[E_1>\log N, \ldots,E_{m^n}>\log N\right],
\label{probHm}
\end{eqnarray} 
where $E_k$'s ($k=1$, $2$, $\ldots$, $m^n$) are respectively the total
energies of the all possible $m^n$ paths. Similarly the cumulative
distribution of the balanced height is related to the distribution of the
maximum energy of the polymer paths via 
\begin{eqnarray} 
{\rm {Prob}}\left[h_N>n\right]={\rm Prob}\left[l_1>1,
  \ldots,\l_{m^n}>1\right] \nonumber \\
= {\rm Prob}\left[E_1<\log N, \ldots,\, E_{m^n}<\log N\right],
\label{probhm}
\end{eqnarray}

\subsection{Statistics of the Height}

Let $P_n(x)={\rm Prob}\left[{\rm min}\{E_1, \ldots, E_{m^n}\}>x
\right]$.  This distribution satisfies the recursion relation
\begin{equation}
P_n(x)=\int_0^{\infty}\ldots\int_0^{\infty}
p(\epsilon_1,\ldots,\epsilon_m)\prod_{i=1}^m
P_{n-1}(x-\epsilon_i)d\epsilon_i,
\label{recur1m}
\end{equation} 
where the joint distribution $p(\epsilon_1,\ldots,\epsilon_m)$ is given
by Eq.~(\ref{jointem}).  The recursion starts with the same initial condition
as in Eq.~(\ref{P0}). The rest of the analysis is exactly same as in the
binary case. Substituting a traveling front solution, $P_n(x)=F(x-vn)$ in Eq.
(\ref{recur1m}) and then linearizing near the tail $y\to -\infty$, we find as
in the binary case, $F(y)\sim 1-e^{\lambda y}$ where the velocity $v$ of the
front is related to $\lambda$ via the dispersion relation 
\begin{equation} 
v(\lambda)=
-{1\over {\lambda}}\log \left[ m\int_0^{\infty}
e^{-\lambda\epsilon}\,\rho(\epsilon)\,d\epsilon\right],
\label{vlm}
\end{equation} 
where the induced distribution $\rho(\epsilon)$ is given by
Eq.~(\ref{rhoem}). The front velocity is then given by the maximum
$v(\lambda^*)$ of the dispersion curve in Eq.~(\ref{vlm}) and is obtained by
solving Eqs.  (\ref{l*}) and (\ref{vlm}). Similarly one can also work out the
logarithmic correction to the front velocity and the asymptotic front
position is given by the same formula in Eq.~(\ref{xn2}), only $\lambda^*$
and $v(\lambda^*)$ are different from the binary case.  Similarly the average
height $\langle H_N\rangle$ for the $m$-ary search tree is also given by the
same formula as in Eq.~(\ref{H_N}), only change is in the dispersion curve
$v(\lambda)$.

Let us now present some specific results.  For the uniform distribution,
$\rho(\epsilon)=(m-1)[1-e^{-\epsilon}]^{m-2}e^{-\epsilon}$ as follows from
Eqs.~(\ref{etar1}) and (\ref{rhoem}).  Substituting this into the dispersion
relation (\ref{vlm}) yields
\begin{equation}
v(\lambda)=-{1\over {\lambda}}\log \left[ m(m-1)B(\lambda+1,m-1)\right],
\label{vlrm1}
\end{equation} 
where $B(m,n)$ is the Beta function. For instance, for $m=3$ the velocity
$v(\lambda)$ has a single maximum at $\lambda^*=3.48985\ldots$ with
$v(\lambda^*)=0.40487\ldots$.  Plugging these in the general formula
(\ref{H_N}) we again arrive at Eq.~(\ref{HN1}) with
$\alpha_0=2.4698\dots$ and $\alpha_1=-1.0616\dots$.  

Consider now the large $m$ limit.  Using asymptotic properties of the Beta
function, one gets 
\begin{equation}
\lambda^* \approx \log m, \qquad
v(\lambda^*)\approx \log (m/\lambda^*).  
\label{largem}
\end{equation} 
Therefore when $m\to\infty$, the average height is given by
Eq.~(\ref{HN1}) with
\begin{equation} 
\alpha_0={1\over {\log (m/\lambda^*)}}, \quad 
\alpha_1= -{3\over {2\lambda^*\log(m/\lambda^*)}}.
\label{HNm1l}
\end{equation} 

Similarly for the distribution (\ref{etar2}), Eqs.~(\ref{rhoem}) and
(\ref{vlm}) lead to the following dispersion relation
\begin{eqnarray*}
v(\lambda)=-{1\over {\lambda}}\log \left[
  m(2m-1)(2m-2)B(\lambda+2,2m-2)\right].
\end{eqnarray*}
For $m=3$, we get the maximum at $\lambda^*=4.17886\ldots$ with
$v(\lambda^*)=0.53235\ldots$.  The average height is given by Eq.~(\ref{HN1})
with $\alpha_0=1.87845\dots$ and $\alpha_1=-0.67427\dots$.  The large $m$
behavior turns out to be exactly the same as in the case of uniform
distribution.  One can work out the large $m$ asymptotics for arbitrary
distribution $\eta(r)$ (see Appendix D) and one gets the same asymptotics
(\ref{largem}) as in the above examples. Therefore, the asymptotic behavior
of $\langle H_N\rangle$ is universal (independent of the details of the
distribution) in the large $m$ limit.

\subsection{Statistics of the Balanced Height}

As in the binary case, we again utilize the distribution $R_n(x)={\rm
  Prob}[{\rm max}\{E_1,E_2,\ldots,E_{m^n}\}<x]$.  This distribution satisfies
the recursion relation
\begin{equation}
R_n(x)=\int_0^{\infty}\ldots
\int_0^{\infty}p(\epsilon_1,\ldots,\epsilon_m)\prod_{i=1}^m
R_{n-1}(x-\epsilon_i)d\epsilon_i,
\label{recurhm}
\end{equation} 
and the same initial and boundary conditions (\ref{R0})--(\ref{Rbound}) as in
the binary case. Plugging a traveling front solution $R_n(x)=G(x-v_1n)$ into
Eq.~(\ref{recurhm}) and linearizing in the tail region $y\to \infty$
according to $G(y)\approx 1-e^{-\mu y}$, we arrive at the dispersion relation
\begin{equation} 
v_1(\mu)= {1\over {\mu}}\log \left[ m\int_0^{\infty}e^{\mu
    \epsilon}\rho(\epsilon)d\epsilon\right],
\label{v1lm}
\end{equation} 
where the induced distribution is given by Eq.~(\ref{rhoem}). The front
velocity is then selected by the minimum $v_1(\mu^*)$ of this dispersion
relation. Proceeding as in the binary case, the asymptotic front position is
given by the same general formula in Eq.~(\ref{xn*2}), the only difference is
that $\mu^*$ and $v_1(\mu^*)$ are different from the binary case. Finally the
average balanced height $\langle h_N\rangle$ for the $m$-ary search trees is
also given by the same general formula in Eq.~(\ref{h_N}), the only
difference being the dispersion relation $v_1(\mu)$.

For the uniform distribution, E[q.~(\ref{etar1})], we reduce Eq.~(\ref{v1lm})
to 
\begin{equation} 
v_1(\mu)={1\over
  {\mu}}\log \left[ m(m-1)B(1-\mu,m-1)\right].
\label{v1lrm}
\end{equation} 
Equation (\ref{v1lrm}) can also be obtained from Eq.~(\ref{vlrm1}) by
changing the sign of $\lambda=-\mu$ as expected. For example, for $m=3$, the
dispersion relation (\ref{v1lrm}) has a unique minimum at
$\mu^*=0.68189\ldots$ where $v_1(\mu^*)=3.90227\ldots$.  Then the
general formula (\ref{h_N}) reduces to Eq.~(\ref{hN1}) with
$\alpha_0'=0.25626\dots$ and $\alpha_1'=0.56371\dots$. 
 
For the distribution (\ref{etar2}), the dispersion relation reads
\begin{eqnarray*} 
v_1(\mu)={1\over
  {\mu}}\log \left[ m(2m-1)(2m-2)B(2-\mu,2m-2)\right].
\end{eqnarray*} 
One has $\mu^*=1.28665\ldots$ and $v_1(\mu^*)=2.62334\ldots$
indicating in the particular case of $m=3$, so in this situation the averaged
balanced height is given by Eq.~(\ref{hN1}) with
$\alpha_0'=0.38119\dots$ and $\alpha_1'=0.44440\dots$. 

One can also work out the large $m$ behavior for arbitrary distribution
$\eta(r)$ (Appendix D).  Unlike the case of the height variable, the large $m$
behavior in the case of balanced height is nonuniversal and depends
explicitly on the small $r$ behavior of the distribution $\eta(r)$. If
$\eta(r)\sim r^{a}$ as $r\to 0$, then (see Appendix D) $\mu^*\approx a+1$ and
$v_1(\mu^*)\approx {{a+2}\over {a+1}}\log m$.  Both these quantities, and
hence the average balanced height, depend on the parameter $a$.  Therefore,
the balanced height remains nonuniversal in the large $m$ limit.

\section{Conclusions}

In this paper we studied the statistics of height and balanced height in the
BST problem by exploiting a two stage mapping ``the BST problem $\to$
fragmentation problem $\to$ the MDP problem" and then using the traveling
front techniques to solve the MDP problem. While the first mapping has been
used previously to obtain exact asymptotic results for RBST problem, the
second mapping allowed us to go beyond random trees and obtain exact
asymptotic results for BST's where the new entries arrive in the tree
according to any arbitrary distribution, not necessarily randomly. The fact
that the traveling wave techniques, used previously in nonlinear physics, can
be used successfully in computer science problems is not just interesting but
it allows us to obtain many new informations such as the shape of the full
distribution of height and not just its moments.  It would be interesting to
apply these techniques to more sophisticated search algorithms in computer
science.

\acknowledgements We thank D.~S.~Dean for useful discussions.  PLK was
partially supported by NSF(DMR9978902).

\appendix
\section{Derivation of the Logarithmic Correction to the Front Position}

In this Appendix, we present a detailed derivation of the logarithmic
correction to the asymptotic front position.  We employ the approach of
Ref.\cite{BD} where such a correction was computed for a reaction diffusion
equation. In the present context, our starting point is the recursion
relation in Eq.~(\ref{recur1m}) for the $m$-ary search trees.  We first
substitute $P_n(x)=1-f_n(x)$ in Eq.~(\ref{recur1m}) and then neglect terms of
order $O(f_n^2)$ in the regime $x\to -\infty$ to get a linear equation
\begin{equation}
f_{n+1}(x)=m\int_0^{\infty}f_n(x-\epsilon)\rho(\epsilon)d\epsilon,
\label{a1}
\end{equation} 
where $\rho(\epsilon)$ is the effective induced distribution
$\rho(\epsilon)$ is given by Eq.~(\ref{rhoem}). Next we assume that for
large $n$ the front position is given by $x_n=vn+c(n)$, where both the
velocity $v$ and the functional form of the correction term $c(n)$ are yet to
be determined. Following Ref. \cite{BD}, we then assume that for large $n$
the solution $f_n(x)$ of Eq.~(\ref{a1}) is given by the scaling form 
\begin{equation}
f_n(x)= n^{\gamma}H\left( {{x-x_n}\over
    {n^{\gamma}}}\right)e^{\lambda(x-x_n)},
\label{a2}
\end{equation} 
where the exponent $\gamma$ and the scaling function $H(y)$ are not yet
known. We only know that $H(y)\to 0$ as $y\to \pm \infty$ (since $0\le
f_n(x)\le 1$ for all $x$). Also, since for large $n$, the prefactor
$n^{\gamma}$ in Eq.~(\ref{a2}) must go away, indicating that $H(y)\sim y$ as
$y\to 0$.

Let us define $z_{n}=(x-x_n)/n^{\gamma}$. Then to leading order for large
$n$, one has $z_{n+1}\approx z_n-{\gamma\over {n}}z_n-vn^{-\gamma}$.
Substituting $z_{n+1}$ in Eq.~(\ref{a2}) and keeping only leading order terms
we get for the left-hand side of Eq.~(\ref{a1}) 
\begin{eqnarray} 
f_{n+1}(x)&\approx & n^{\gamma}
e^{\lambda (x-x_n-v)}[ H(z)-{{vH'(z)}\over {n^{\gamma}} } -{\gamma\over
  {n}}zH'(z)
\nonumber \\
&+& \left({\gamma\over {n}}-\lambda {dc\over dn}(n)\right)H(z) +{v^2\over
  {2n^{2\gamma}}}H''(z)].
\label{a3}
\end{eqnarray}
In the above equation, we used the shorthand notations $z_n=z$,
$H'(z)=dH/dz$, and $H''(z)=d^2H/dz^2$. 

Similarly, inserting Eq.~(\ref{a2}) into the right-hand side of
Eq.~(\ref{a1}), expanding $H[(x-x_n-\epsilon)n^{-\gamma}]$ in Taylor series
in $\epsilon e^{-\gamma}$, and keeping only leading order terms, we find the
right-hand side of Eq.~(\ref{a1})
\begin{eqnarray}
f_{n+1}(x)&\approx& mn^{\gamma}e^{\lambda(x-x_n)}[\mu_0 H(z)-{\mu_1\over
  {n^{\gamma}}}H'(z)
\nonumber \\
&+& {\mu_2\over {2n^{2\gamma}}}H''(z)],
\label{a4}
\end{eqnarray} 
where $\mu_k=\int_0^{\infty} {\epsilon}^k e^{-\lambda
  \epsilon}\rho(\epsilon)d\epsilon $. Comparing the left-hand side given by
Eq.~(\ref{a3}) and the right-hand side given by Eq.~(\ref{a4}), we recover,
to leading order for large $n$, the dispersion relation
\begin{equation} 
e^{-\lambda v}=m\int_0^{\infty}e^{-\lambda \epsilon}\rho(\epsilon)d\epsilon.
\label{a5}
\end{equation} 
As argued before, the front will choose the maximum velocity $v(\lambda^*)$
of the dispersion relation (\ref{a5}). At $\lambda=\lambda^*$,
$v'(\lambda^*)=0$. Differentiating Eq.~(\ref{a5}) with respect to $\lambda$
we obtain $v(\lambda^*)\exp[-\lambda^* v(\lambda^*)]=m\mu_1$.  Using this
in Eq.~(\ref{a3}) shows that the term of order $n^{-\gamma}$ in
Eq.~(\ref{a3}) cancels the corresponding term on the right-hand side in
Eq.~(\ref{a4}).  To ensure that remaining terms are of the same order, we
must have $\gamma=1/2$ and $dc/dn=b/n$. The latter equation gives $c(n)=b
\log n$, where $b$ is still undetermined.  Employing these choices for
$\gamma$ and $c(n)$ and equating Eqs.~(\ref{a3}) and (\ref{a4}), we obtain
\begin{eqnarray*} 
\left(v^2-me^{\lambda^* v}\mu_2\right)H''(z)-zH'(z)+(1-2b\lambda^*)H(z)=0,
\end{eqnarray*} 
where $v=v(\lambda^*)$. This equation can be further simplified as follows.
Differentiating Eq.~(\ref{a5}) twice with respect to $\lambda$ and using
$v'(\lambda^*)=0$ we get an additional relation,
$v^2(\lambda^*)-m\mu_2\exp[\lambda^*v(\lambda^*)] =\lambda^*v''(\lambda^*)$.
By inserting this into the above equation we finally arrive at the eigenvalue
equation
\begin{equation} 
-\lambda^*v''(\lambda^*)H''(z)+zH'(z)+(2b\lambda^*-1)H(z)=0.
\label{a7}
\end{equation} 
Note that $v(\lambda)$ has a maximum at $\lambda=\lambda^*$ indicating
$v''(\lambda^*)<0$. Rescaling $z=\sqrt{-\lambda^*v''(\lambda^*)}\,\zeta$, we
find that the solution of Eq.~(\ref{a7}) that vanishes at $\zeta\to \infty$ is
given by $H(\zeta)=B e^{-\zeta^2/4}D_{2b\lambda^*-2}(\zeta)$, where $B$ is a
constant and $D_p(\zeta)$ is the parabolic cylinder function of index $p$.
The condition that $H(\zeta)\sim \zeta$ as $\zeta\to 0$ enforces the choice
of the index $p=2b\lambda^*-2=1$ indicating $b=3/{2\lambda^*}$. Note that the
above solution describes precisely the wave function of the first excited
state of a quantum harmonic oscillator and the factor $3/2$ is the
corresponding energy eigenvalue.  Finally, the leading asymptotic behavior of
the front position is given by
\begin{equation} 
x_n = v(\lambda^*)n + {3\over {2\lambda^*}}\log n.
\label{a8}
\end{equation} 

A similar calculation can be carried out for the balanced height where one
finds a dispersion relation $v_1(\mu)$ as given by Eq.~(\ref{v1lm}) and front
position is given by 
\begin{equation} 
{x_n}^* = v_1(\mu^*) - {3\over {2\mu^*}}\log n,
\label{a9}
\end{equation} 
where $\mu^*$ denotes the point where $v_1(\mu)$ has its unique minimum.

\section{Asymptotic Behavior of the Cumulative Height Distribution}

In this Appendix, we derive the large $y$ behavior of the cumulative height
distribution $F(y)$.  The function $F(y)$ is the solution of the boundary
value problem (\ref{Fy})--(\ref{Fbound}).  We already know that $1-F(y)\sim
e^{\lambda^* y}$ as $y\to -\infty$, where $\lambda^*$ denotes the value of
$\lambda$ where the dispersion curve $v(\lambda)$ in Eq.~(\ref{vl}) has its
maximum. In order to derive the asymptotic behavior of $F(y)$ in the other
limit $y\to \infty$, we first recast the integral equation (\ref{Fy}) in a
slightly different form.  Let us first define the cumulative distribution
function
\begin{equation} 
Y(\epsilon,\epsilon')=\int_0^{\epsilon}\int_0^{\epsilon'}
p(x_1,x_2)\,dx_1\, dx_2,
\label{b1}
\end{equation} 
where the joint distribution $p(x_1,x_2)$ is given by Eq.~(\ref{rhoe}).
Writing $p(\epsilon,\epsilon')= {\partial^2 Y}/{{\partial \epsilon}{\partial
    \epsilon'}}$ on the right-hand side of Eq.~(\ref{Fy}) and performing the
integrations by part (first over $\epsilon$ and then over $\epsilon'$), we
finally arrive at the following equation
\begin{eqnarray}
F(y-v)=&&\int_0^{\infty}\int_0^{\infty} F'(y-\epsilon)\,F'(y-\epsilon')\,
Y(\epsilon,\epsilon')\,d\epsilon\, d\epsilon'\nonumber \\
&&+2\int_0^{\infty} F(y-\epsilon)\,\rho(\epsilon)\,d\epsilon - 1,
\label{b2}
\end{eqnarray} 
where $F'(y)=dF/dy$ and we have used the boundary conditions of $F(y)$.  Note
that due to the concentration of measure, $F(y)$ has roughly the
shape of the step function, $F(y)\approx \theta(-y)$ with the front located
at $y=0$.  Thus the derivative roughly behaves as a negative delta function,
$F'(y)\approx -\delta (y)$.  First reconsider the limit $y\to -\infty$. In
this limit, the arguments of the functions $F'(y)$ inside the integrands in
the first term on the right-hand side in Eq.~(\ref{b2}) are always very large
and negative, indicating that the contribution from this term is negligible
as $y\to -\infty$.  Neglecting the first term, one finds that the resulting
linear equation admits the exponential solution $1-F(y)\sim e^{\lambda y}$
where $v$ depends on $\lambda$ through the dispersion relation in
Eq.~(\ref{vl}). Thus one recovers the correct result in the $y\to -\infty$
limit.

Turn now to the complementary limit $y\to \infty$.  Then the arguments of
$F'(y)$ inside the integrands of the first term on the right-hand side of
Eq.~(\ref{b2}) can be close to zero to pick up a substantial contribution.
For large $y$, one can approximate $F'(y)\approx -\delta (y)$ inside the
integrands on the right-hand side of Eq.~(\ref{b2}) and one then gets
\begin{eqnarray}
F(y-v)\approx &-&1+Y(y,y) \nonumber \\
&+& 2\int_0^{\infty}F(y-\epsilon)\,\rho(\epsilon)\,d\epsilon.
\label{b3}
\end{eqnarray} 
$Y(y,y)\to 1$ and $F(y)\to 0$ as $y\to \infty$. To find the asymptotics of
$F(y)$ we differentiate Eq.~(\ref{b3}) with respect to $y$ and use
$F'(y)\approx -\delta(y)$ in the second term. This gives 
\begin{eqnarray} 
F'(y-v) \approx -2\rho(y) + 2{{\partial Y(y,y_2)}\over
  {\partial y_2}}\Big|_{y_2=y}.
\label{b4}
\end{eqnarray} 
Using the definitions in Eqs.~(\ref{b1}) and (\ref{rhoe}) we find 
\begin{eqnarray*}  
{{\partial Y}\over {\partial y_2}}
=-e^{-y_2}\,\phi(e^{-y_2})\,\phi(1-e^{-y_2})\,\theta(e^{-y_1}+e^{-y_2}-1).
\end{eqnarray*} 
When $y_1=y_2$ is large, the argument of the step function in the above
equation is always negative, indicating that one can neglect the second term
on the right-hand side of Eq.~(\ref{b4}). This gives $F'(y) \approx
-2\rho(y+v)$. Hence the desired large $y$ behavior of $F(y)$ is
given by
\begin{equation} 
F(y) \approx 2\int_y^{\infty}\rho(y'+v)\,dy',
\label{b6}
\end{equation} 
where $v=v(\lambda^*)$ is the maximum velocity associated with the
dispersion relation (\ref{vl}).

Note that the constraint $e^{-\epsilon}+e^{-\epsilon'}=1$ does not modify the
form of the dispersion curve when compared to the unconstrained conventional
DP problem [the only difference is that one has to first find the effective
single point energy distribution $\rho(\epsilon)$ in the constrained case
from Eq.~(\ref{rhoe})].  However, the above constraint does modify the large
$y$ behavior of the cumulative distribution $F(y)$. For example,
Eq.~(\ref{b4}) is valid for the unconstrained problem as well.  However in
the unconstrained case, $Y(y,y)= [\int_0^{y}\rho(\epsilon)\, d\epsilon]^2$.
In that case one finds after taking the derivative, $F'(y) \approx
-2\rho(y+v)\int_y^{\infty} \rho(\epsilon)d\epsilon$ indicating that for large
$y$
\begin{equation} 
F(y)|_{\rm unconstrained}\approx 2\int_y^{\infty}dy'
\rho(y'+v)\int_{y'}^{\infty}\rho(\epsilon)d\epsilon.
\label{b7}
\end{equation} 
For example, for the RBST where $\rho(\epsilon)=e^{-\epsilon}$, the large $y$
asymptotics are $F(y)\sim e^{-y}$ (constrained case) and $F(y) \sim e^{-2y}$
(unconstrained case).
 
\section{ Derivation of the Induced Distribution}

In this Appendix we derive the induced distribution $\eta(r)$ [see
Eq.~(\ref{etar2})] starting from the joint distribution
\begin{equation} 
{\rm Prob}[r_1,\ldots,r_m]=A_m \,\delta\left(\sum_{i=1}^m
  r_i-1\right)\prod_{i=1}^m r_i.
\label{c1}
\end{equation} 
The constant $A_m$ in the above equation has to be chosen such that the joint
distribution is normalized. The induced distribution $\eta(r)$ is obtained by
fixing one of the fractions, say the first one, to the value $r$ and then
integrating over all other fractions. Thus by definition
\begin{equation} 
\eta(r)= A_m r
\int_0^{1}\ldots\int_0^{1}\delta\left(\sum_{i=2}^m r_i
  +r-1\right)\prod_{i=2}^m r_i dr_i.
\label{c2}
\end{equation}

Note that $r_i$'s denote the lengths of $m$ intervals with the total length
equal to unity.  Let us define a set of new variables, $x_2=r+r_2$,
$x_3=x_2+r_3$, $\ldots$, $x_{m-1}=x_{m-2}+r_{m-1}$. Here $x_i$'s denote the
points separating adjacent intervals. Clearly then $x_{m-1}=1-r_m$ since the
total length is unity. With these change of variables the integral in
Eq.~(\ref{c2}) becomes
\begin{equation} 
\eta(r)=A_m r \zeta_m(r),
\label{c3}
\end{equation} 
where $\zeta_m(r)$ is given by 
\begin{eqnarray}
\zeta_m(r)&=&\int_r^{1}(x_2-r)dx_2\int_{x_2}^1
(x_3-x_2)dx_3\dots \nonumber \\
&\dots&\int_{x_{m-2}}^1(x_{m-1}-x_{m-2})(1-x_{m-1})dx_{m-1}.
\label{c4}
\end{eqnarray} 
Thus $\zeta_m(r)$ satisfies the recursion relation 
\begin{equation}
\zeta_m(r)=\int_r^1 (x_2-r)\zeta_{m-1}(x_2)dx_2.
\label{c5}
\end{equation} 
One directly computes $\zeta_2(r)=1-r$ and
$\zeta_3(r)=(1-r)^3/6$ which suggests to seek a solution in the form 
$\zeta_m(r)=B_m(1-r)^{2m-3}$.  Plugging the above expression in recursion 
(\ref{c5}) yields
\begin{equation}
B_m={{B_{m-1}}\over {(2m-3)(2m-4)}},
\label{c6}
\end{equation} 
which is iterated to give $B_m=1/(2m-3)!$.  Thus we obtain $\eta(r)=A_m
r(1-r)^{2m-3}/(2m-3)!$. The normalization condition $\int_0^{1}\eta(r)dr=1$
then gives $A_m=\Gamma(2m)$ where $\Gamma(x)$ is the Gamma function.
Therefore
\begin{equation} 
\eta(r)=
(2m-1)(2m-2)r(1-r)^{2m-3},
\label{c7}
\end{equation} 
which is valid for all $m\ge 2$.

\section{Large $m$ Results for Arbitrary Distribution}

In this appendix we derive the large $m$ behavior of $\langle H_N\rangle$ and
$\langle h_N\rangle$ for $m$-ary search trees with arbitrary distribution
$\eta(r)$.  We start with the height variable and write the
dispersion relation 
\begin{eqnarray}
e^{-\lambda v}&=&m\int_0^{\infty} 
e^{-\lambda \epsilon}\,\rho(\epsilon)\,d\epsilon \nonumber \\
&=& m \int_0^{1}r^{\lambda}\, \eta(r)\, dr.
\label{d1}
\end{eqnarray} 
The constraint $\sum r_i=1$ leads to $\int_0^{1} r\eta(r)dr=1/m$. Thus for
large $m$, a generic distribution $\eta(r)$ will have be concentrated near
$r=0$.  Consider a class of distributions which behave as $\eta(r)\approx C_m
r^a e^{-b_m r}$ near the origin. For example, $C_m=m-1$, $a=0$, and $b_m=m-2$
for the uniform distribution (\ref{etar1}).  Similarly, $C_m=(2m-1)(2m-2)$,
$a=1$ and $b_m=2m-3$ for the distribution (\ref{etar2}).  These two examples
suggest that $C_m\sim m^{a+1}$ and $b_m\sim m$.  Making use of the
constraints $\int_0^{1}\eta(r)dr=1$ and $\int_0^{1}r\eta(r)dr=1/m$ one indeed
confirms the above asymptotics: $b_m\approx (a+1)m$ and
$C_m\approx b_m^{a+1}/\Gamma(a+1)$.

We now consider the integral in Eq.~(\ref{d1}).  Substituting the small $r$
behavior of $\eta(r)$, performing the integral, and using the Stirling
formula one gets
\begin{equation}
\left(b_m e^{-v}\right)^\lambda \approx {m\over \Gamma(a+1)}\, 
\sqrt{2\pi(\lambda+a)}\,\left({\lambda+a\over  e}\right)^{\lambda+a}.
\label{d2}
\end{equation} 
Taking the logarithm, differentiating with respect to $\lambda$, and setting
$v'(\lambda^*)=0$ we determine $\lambda^*$ and $v(\lambda^*)$.  The leading
contributions are given by Eq.~(\ref{largem}).  Therefore the large $m$
behavior of $\langle H_n\rangle$ is indeed universal.

We now turn to the large $m$ behavior of the average balanced height $\langle
h_N\rangle$.  In this case, the appropriate dispersion relation is given 
by Eq.~(\ref{v1lm}): 
\begin{eqnarray}
e^{\mu v_1}&=&m\int_0^{\infty}e^{\mu \epsilon}
\rho(\epsilon)d\epsilon \nonumber \\
&=& m \int_0^{1}r^{-\mu}\eta(r)dr.
\label{d5}
\end{eqnarray} 
Substituting the small $r$ behavior, $\eta(r)\approx C_m
r^{a}e^{-b_mr}$, and performing the integral we obtain
\begin{equation} 
\left(b_m^{-1} e^{v_1}\right)^\mu \approx {m\over \Gamma(a+1)}\, 
\Gamma(a+1-\mu).
\label{d6}
\end{equation} 
We will see that in the large $m$ limit, $\mu^* \to a+1$. Hence we write
$\mu\to a+1-\delta$, assume that $\delta\ll 1$, plug these in
Eq.~(\ref{d6}) and take the logarithm to obtain
\begin{equation} 
(a+1-\delta)(v_1-\log b_m)\approx \log{m\over\Gamma(a+1)}  -\log \delta.
\label{d7}
\end{equation} 
Differentiating Eq.~(\ref{d7}) with respect to $\delta$ and setting
$v'(\delta^*)=0$ yields:
\begin{eqnarray*}
\mu^*&=&a+1-{a+1\over \log m}+\ldots,\\
v(\lambda^*)&=&{a+2\over a+1}\,\log m+\ldots.
\end{eqnarray*} 
The parameter $a$ appears in the leading order even in the large $m$ limit.
Consequently, $\langle h_N\rangle$ also depends on $a$ and thence the
balanced height is not universal in the large $m$ limit.

\end{multicols}
\end{document}